\pdfoutput=1 
\documentclass[aps,twocolumn,prl,showpacs,amsmath,amssymb,10pt]{revtex4-1}
\usepackage{graphicx}
\usepackage{scrextend}
\usepackage{manfnt,pifont}

\usepackage{url}

\begin{document}

%\preprint{}

\title{Large Spin Perturbation Theory}

\author{Luis F. Alday}

\affiliation{Mathematical Institute, University of Oxford,\\ Andrew Wiles Building, Radcliffe Observatory Quarter\\ Woodstock Road, Oxford, OX2 6GG, UK}

\date{\today}

\begin{abstract}
\noindent 
We consider conformal field theories around points of large twist degeneracy. Examples of this are theories with weakly broken higher spin symmetry and perturbations around generalised free fields. At the degenerate point we introduce twist conformal blocks. These are eigenfunctions of certain quartic operators and encode the contribution, to a given four-point correlator, of the whole tower of intermediate operators with a given twist. As we perturb around the degenerate point, the twist degeneracy is lifted. In many situations this breaking is controlled by inverse powers of the spin. In such cases the twist conformal blocks can be decomposed into a sequence of functions which we systematically construct. Decomposing the four-point correlator in this basis  turns crossing symmetry into an algebraic problem. Our method can be applied to a wide spectrum of conformal field theories in any number of dimensions and at any order in the breaking parameter. As an example, we compute the spectrum of various theories around generalised free fields. For theories with higher spin symmetry we discuss the relation between twist conformal blocks and higher spin conformal blocks.

\end{abstract}

\maketitle

\section{Introduction and summary}

A wide spectrum of physical phenomena, from boiling water to the horizon of certain black holes, is described by conformal field theories (CFT). A CFT is characterised by a set of local primary operators ${\cal O}_{\Delta,\ell}(x)$ labelled by their scaling dimension $\Delta$ and a $SO(d)$ representation $\ell$. These operators satisfy an algebra, whose structure constants are denoted OPE coefficients. The spectrum of scaling dimensions $\{\Delta_i,\ell_i\}$ and OPE coefficients constitutes the {\it CFT data}. All correlators of local operators can in principle by written in terms of the CFT data. Explicit determination of this data for a specific CFT leads to a better understanding of all the physical phenomena which this CFT describes. 

Many analytic results can be obtained in regimes where there is a small parameter. These computations, in essence perturbative, include $\epsilon-$expansions, large $N$ computations and perturbative computations. The idea of the conformal bootstrap \cite{Ferrara:1973yt,Polyakov:1974gs} is to use instead associativity of the operator algebra to constraint the CFT data.  In two dimensions conformal symmetry is enhanced to the infinite dimensional Virasoro symmetry and the conformal bootstrap led to the full solution of a vast family of theories, called minimal models. In higher dimensions progress was harder, but following \cite{Rattazzi:2008pe}, the conformal bootstrap has led to vast numerical results for CFT in various dimensions and motivated the search for analytic results without using Lagrangian/perturbative methods. A set of analytic results along these lines involves the large spin sector of the theory. This was studied in \cite{Alday:2007mf} and then systematically, from the point of view of crossing symmetry, in \cite{Komargodski:2012ek,Fitzpatrick:2012yx} for generic CFT and in \cite{Alday:2013cwa,Alday:2015ota} for weakly coupled CFT. See also \cite{Kaviraj:2015cxa,Kaviraj:2015xsa,Alday:2015eya,Alday:2015ewa} for interesting extensions. A remarkable conclusion of all these results is that the large spin sector of a CFT is universal and behaves as in a free theory.  Another set of analytic results from the conformal bootstrap/crossing symmetry perspective involves expansions around small parameters. These include results for large $N$ gauge theories \cite{Heemskerk:2009pn} and results in the $\epsilon-$expansion \cite{Gopakumar:2016wkt}, among others. 

The aim of the present letter is to connect these developments. We consider a CFT around a point of large twist degeneracy. At the degenerate point we assume the spectrum of  a given twist is unbounded in the spin. It is convenient to introduce twist conformal blocks (TCB) $H^{(0)}_\tau(u,v)$ in which four-point correlators decompose
\begin{equation}
{\cal G}^{(0)}(u,v) = \sum_\tau H^{(0)}_\tau(u,v)
\end{equation}
As we perturb around the degenerate point, operators acquire anomalous dimensions and the degeneracy in the twist is lifted. In many situations this breaking is proportional to inverse powers of the spin and the TCB can be further decomposed into a sequence of functions $H^{(\rho)}_\tau(u,v)$, where $\rho$ measures the departure from the degenerate value. The correlator can now be decomposed  
\begin{equation}
{\cal G}(u,v) = \sum_{\tau,\rho} H^{(\rho)}_\tau(u,v)
\end{equation}
A great advantage of this decomposition is that in the Lorentzian regime the functions $H^{(\rho)}_\tau(u,v)$ have well understood behaviour around $u,v \sim 0$ and this is such that the crossing equation becomes essentially algebraic! Our method can be applied to vast families of CFT, {\it e.g.} theories with weakly broken HS symmetry in any number of dimensions, large $N$ theories, etc. As an example, we compute the spectrum of various theories around generalised free fields, and comment on other cases. 

\section{Degenerate point}
Consider the four-point correlator of identical scalar operators in a generic CFT in $d-$dimensional Minkoski space
\begin{equation}
\langle \phi(x_1) \phi(x_2) \phi(x_3) \phi(x_4) \rangle = \frac{{\cal G}(u,v)}{x_{12}^{2 \Delta_\phi} x_{34}^{2\Delta_\phi}}   
 \end{equation}
 where we have introduced cross-ratios $u=\frac{x_{12}^2x_{34}^2}{x_{13}^2x_{24}^2}$ and $v=\frac{x_{14}^2x_{23}^2}{x_{13}^2x_{24}^2}$. Crossing symmetry implies
\begin{equation}
v^{\Delta_\phi} {\cal G}(u,v) = u^{\Delta_\phi} {\cal G}(v,u).
 \end{equation}
 The four point correlator admits a decomposition in conformal partial waves
 \begin{equation}
\label{CPW}
{\cal G}(u,v) =1+ \sum_{\Delta,\ell} a_{\Delta,\ell} G_{\Delta,\ell}(u,v) 
 \end{equation}
where $G_{\Delta,\ell}(u,v)$ denote the corresponding conformal blocks. We will also use the notation $G_{\Delta,\ell}(u,v) = u^{\frac{\tau}{2}} g_{\tau,\ell}(u,v)$, where $\tau=\Delta-\ell$ denotes the twist. This notation makes manifest the small $u$ behaviour of conformal blocks. 

We will further assume the CFT under consideration has a small parameter $g$, which measures perturbations around a degenerate point. At $g=0$ the spectrum of twists is assumed to be highly degenerate, such that for a given twist $\tau$ there is an infinite tower of operators of (unbounded) spin $\ell$. We then consider the following functions

\begin{equation}
\sum_{\ell} a_{\tau,\ell}^{(0)} u^{\tau/2} g_{\tau,\ell}(u,v) = H^{(0)}_{\tau}(u,v)
\end{equation}
where $a_{\tau,\ell}^{(0)}$ are the square of the OPE coefficients at $g=0$. $H^{(0)}_{\tau}(u,v)$ encodes the contribution from a given twist, and all spins, to the correlator in the degenerate point. We call these functions twist conformal blocks (TCB). The four point correlator can be written as

\begin{equation}
 {\cal G}^{(0)}(u,v) = \sum_{\tau} H^{(0)}_{\tau}(u,v)
\end{equation}
which expresses the decomposition of the correlator at $g=0$ into TCB. 

\subsection{Twist conformal blocks}
In order to construct the functions $H^{(0)}_{\tau}(u,v)$ we need to understand their properties. First, from the small $u$ behaviour of conformal blocks we can deduce

\begin{equation}
H^{(0)}_{\tau}(u,v) \sim u^{\tau/2} ~~~~\text{at small $u$}
\end{equation}
To understand the small $v$ limit is more subtle, since the definition of TCB involves an infinite sum over the spin, which can enhance the divergent behaviour of a single conformal block. This behaviour can be determined following \cite{Alday:2007mf,Komargodski:2012ek,Fitzpatrick:2012yx,Alday:2013cwa}. For the simplest scenario
\begin{equation}
\label{smallv}
H^{(0)}_{\tau}(u,v) \sim \frac{1}{v^{\Delta_\phi}} ~~~~\text{at small $v$}
\end{equation}
In general we could have considered $H^{(0)}_{\tau}(u,v) \sim v^{-\alpha}$, but the particular boundary condition (\ref{smallv}) will be relevant for generalised free fields, to be studied below. Note that this is also consistent with the presence of the identity operator together with crossing symmetry and with the behaviour for the total sum, which should be 
\begin{equation}
\sum_{\tau} H^{(0)}_{\tau}(u,v) = \left(\frac{u}{v} \right)^{\Delta_\phi} + \cdots
\end{equation}
Another important property of TCB in that they are eigenfunctions of a quartic differential operator. This can be seen by recalling that conformal blocks are eigenfunctions of quadratic and quartic Casimir operators\cite{Dolan:2011dv,Hogervorst:2013kva}. More precisely
\begin{eqnarray}
{\cal D}_2 G_{\Delta,\ell}(u,v) &=& \lambda_2 G_{\Delta,\ell}(u,v),\\
{\cal D}_4 G_{\Delta,\ell}(u,v) &=& \lambda_4 G_{\Delta,\ell}(u,v)
\end{eqnarray}
where 
\begin{eqnarray}
\lambda_2 &=& \frac{1}{2} \left( \ell(\ell+d-2)+(\tau+\ell)(\tau+\ell-d) \right)\\
\lambda_4 &=&\ell(\ell+d-2)(\tau+\ell-1)(\tau+\ell-d+1)
\end{eqnarray}
and the Casimir operators are given by
\begin{eqnarray}
{\cal D}_2 &=&D+\bar D + (d-2) \frac{z \bar z}{z-\bar z}\left( (1-z) \partial -(1-\bar z) \bar \partial \right) \\
{\cal D}_4 &=& \left( \frac{z \bar z}{z-\bar z}\right)^{d-2} (D-\bar D)\left( \frac{z \bar z}{z-\bar z}\right)^{2-d}(D-\bar D)
\end{eqnarray}
where we have introduced variables $u = z \bar z,v=(1-z)(1-\bar z)$ and $D=(1-z)z^2 \partial^2 - z^2 \partial$. Given the explicit form of the eigenvalues $\lambda_2$ and $\lambda_4$ we can consider the following combination
\begin{equation}
{\cal H}_\tau = {\cal D}_4 - {\cal D}_2^2+(d^2-d (2 \tau +3)+\tau ^2+2 \tau +2) {\cal D}_2
\end{equation}
of which TCB are eigenfunctions
\begin{equation}
\label{Tequation}
{\cal H}_\tau H^{(0)}_{\tau}(u,v) = \lambda H^{(0)}_{\tau}(u,v)
\end{equation}
with the following eigenvalue
\begin{equation}
\lambda=\frac{\tau}{4}\left(d^2 (5 \tau +6)-2 d \left(2 \tau ^2+5 \tau +2\right)+\tau  (\tau +2)^2 -2 d^3\right)
\end{equation}
Note that in obtaining the eigenvalue equation, and the boundary conditions, little information was needed about the explicit form of the conformal blocks, or the OPE coefficients at $g=0$. The eigenvalue equation (\ref{Tequation}), together with the boundary conditions and some information about the theory at $g=0$ suffices to fix the form of the TCB. Let us see how this works in more detail. Around $v=0$ we expect the following structure
\begin{equation}
H^{(0)}_{\tau}(u,v) = \frac{1}{v^{\Delta_\phi}} \left( h_\tau^{(0)}(u)+h_\tau^{(1)}(u) v + \cdots \right)
\end{equation}
Plugging this expansion into the eigenvalue equation (\ref{Tequation}) we obtain a sequence of second order differential equations for the functions $h_\tau^{(i)}(u)$ which can be solved iteratively. For instance, the equation for $h_\tau^{(0)}(u)$ leads to two independent solutions, which at small $u$ behave as $u^{\tau/2}$ and $u^{d-1-\tau/2}$. Keeping the one with the correct behaviour we obtain
\begin{equation}
h_\tau^{(0)}(u) = c_0 (1-u)^{1-\frac{d}{2}+\Delta_\phi} u^{\frac{\tau}{2}} F_\frac{2+\tau-d}{2}(u)
\end{equation}
where $F_{\beta}(u)=~_2F_1(\beta,\beta,2\beta;u)$ is the standard hypergeometric function. Plugging this into the next equation we obtain a second order equation for $h_\tau^{(1)}(u)$. Requiring the correct small $u$ behaviour leave us with another arbitrary coefficient, let's say $c_1$, and so on. The situation is particularly simple in $d=2$. The eigenvalue equation can be solved to all orders and the solution takes the factorised form
\begin{equation}
H^{(0)}_{\tau}(z,\bar z) = \bar H^{(0)}_\tau(\bar z) z^{\frac{\tau}{2}}F_\frac{\tau}{2}(z)
\end{equation}
where $\bar H^{(0)}_\tau(\bar z)  \sim (1-\bar z)^{-\Delta_\phi}$ for $\bar z$ close to one. In order to understand how to fix $H^{(0)}_{\tau}(z,\bar z)$ completely, it is instructive to look at a specific example. 
\subsection{Example}
Let us consider generalised free fields. In this case
\begin{equation}
G^{(0)}(u,v)  = 1+u^{\Delta_\phi}+ \left(\frac{u}{v}\right)^{\Delta_\phi}
\end{equation}
which indeed satisfies crossing symmetry. The spectrum of intermediate operators is given by double trace operators with twist
\begin{equation}
\tau_{n}=2\Delta_\phi+2n
\end{equation}
The OPE coefficients at zeroth order can be found for instance in \cite{Heemskerk:2009pn}, but their explicit form will not be used here. Let us now consider the decomposition in TCB
\begin{equation}
G^{(0)}(u,v)  = \sum_{n=0}^\infty H^{(0)}_{\tau_n}(u,v)
\end{equation}
The functions $H^{(0)}_{\tau_n}(u,v)$ can be fixed as follows. Let us consider the small $u,v$ expansion of each of them

\begin{eqnarray}
H^{(0)}_{\tau_0}(u,v) &=& \frac{ u^{\Delta_\phi} \left(c^{(0)}_0 + \cdots \right)}{v^{\Delta_\phi}}+\frac{  u^{\Delta_\phi} \left(c^{(1)}_0 + \cdots \right)}{v^{\Delta_\phi-1}}+ \cdots \\
H^{(0)}_{\tau_1}(u,v) &=& \frac{ u^{\Delta_\phi+1} \left(c^{(0)}_1 + \cdots \right)}{v^{\Delta_\phi}}+\frac{  u^{\Delta_\phi+1} \left(c^{(1)}_1 + \cdots \right)}{v^{\Delta_\phi-1}} + \cdots \nonumber
\end{eqnarray}
and so on. As discussed above, the eigenvalue equation for the TCB fixes the rest of the coefficients once the leading coefficients $c^{(0)}_n, c^{(1)}_n, \cdots$ are fixed.  Let us look into the leading term $H_{\tau_0}(u,v)$. The explicit form of the divergence of ${\cal G}^{(0)}(u,v)$ as $\bar z \to 1$ leads to

\begin{equation}
c^{(0)}_0 =1,~~~c^{(1)}_0 =c^{(2)}_0=\cdots=0 
\end{equation}
Which fixes completely $H_{\tau_0}(u,v)$. This function contains sub-leading terms in $u$. Canceling these sub-leading terms will then fix all $c^{(i)}_1$, and so on. In carrying out this procedure it is convenient to think of $\Delta_\phi$ as arbitrarily large, and then analytically continue in this parameter. For instance, carrying out this procedure for $d=2$ we find the following closed expression for the TCB
\begin{equation}
\label{TCBd2}
H_{\tau}^{(0)}(u,v) = c_\tau \left( \frac{\bar z}{1-\bar z}\right)^{\Delta_\phi} z^{\tau/2}F_{\tau/2}(z)
\end{equation}
for $\tau=2\Delta_\phi+2n$ and
\begin{equation}
c_\tau = \frac{\sqrt{\pi } 2^{2-\tau } \Gamma \left(\frac{\tau }{2}\right) \Gamma \left(\Delta_\phi+\frac{\tau }{2}-1\right)}{\Gamma (\Delta_\phi)^2 \Gamma \left(\frac{\tau -1}{2}\right) \Gamma \left(\frac{\tau }{2}+1-\Delta_\phi\right)}
\end{equation}
The expression (\ref{TCBd2}) contains all terms around $z=0,\bar z=1$ and for large enough $\Delta_\phi$. The full expression for the TCB should also have the symmetry $u \to u/v,v \to 1/v$ which corresponds to the exchange of operators at $x_1$ and $x_2$. This symmetry is not visible in the regime we have considered, but the full TCB can be obtained as follows. In terms of holomorphic variables the symmetry corresponds to $z \to \frac{z}{z-1}$. Under this 
\begin{equation}
\left( \frac{\bar z}{1-\bar z}\right)^{\Delta_\phi} z^{\tau/2}F_{\tau/2}(z) \to (-1)^{\frac{\tau}{2}} \bar z^{\Delta_\phi} z^{\tau/2}F_{\tau/2}(z)
\end{equation}
The factor $(-1)^{\frac{\tau}{2}} \sim (-1)^n$ leads to the factor $z^{\Delta_\phi}$ when performing the sum over $n$. The full TCB has these two pieces and the full symmetry is recovered. At any rate, in what follows we will only focus in the divergent part of the TCB, and this will suffice to obtain non-trivial information about the spectrum of the theory. 

\section{Breaking the twist degeneracy}

We will now consider the effect of turning on the breaking parameter $g$. As we do this the spectrum and OPE coefficients will acquire a small anomalous dimension. In order to deal with this problem we introduce a shifted Casimir operator ${\cal C}$, given by
\begin{equation}
{\cal C}_\tau ={\cal D}_2 +\frac{1}{4} \tau(2d-\tau-2)
\end{equation}
Conformal blocks are eigenfunctions of this operator, with the conformal spin  
\begin{equation}
J^2_{\tau,\ell} = \frac{1}{4}(2\ell+\tau)(2\ell+\tau-2),
\end{equation}
as eigenvalue. We will assume corrections to the spectrum and OPE coefficients admit an expansion around large spin $\ell$. More precisely, we assume expansions of the form
\begin{eqnarray}
\tau_\ell &=& \tau + g \sum_\rho \frac{c^{(\rho)}_{\tau}}{J^{\rho}_{\tau,\ell}} \\
a_{\tau,\ell} &=&a_{\tau,\ell}^{(0)}\left( 1+ g \sum_\rho \frac{d^{(\rho)}_{\tau}}{J^{\rho}_{\tau,\ell}} \right)
\end{eqnarray}
We will analyse the crossing equation for this problem. As we will see, this can be converted into an algebraic problem 
by defining the following set of functions:
\begin{equation}
\sum_{\ell} a_{\tau,\ell}^{(0)} \frac{u^{\tau/2}}{J_{\tau,\ell}^{2m}} g_{\tau,\ell}(u,v) = H^{(m)}_{\tau}(u,v)
\end{equation}
$H^{(0)}_{\tau}(u,v)$ coincides with the TCB introduced above, while $m$ "measures" the departure from the degenerate point. The functions $H^{(m)}_{\tau}(u,v)$ satisfy the following recursion relations
\begin{equation}
\label{recurrence}
{\cal C} H^{(m+1)}_{\tau}(u,v)= H^{(m)}_{\tau}(u,v)
\end{equation}
While conformal blocks are eigenfunctions of ${\cal C}$, TCB are not. Instead, the quadratic operator ${\cal C}$ move us along the sequence of functions $H^{(m)}_{\tau}(u,v)$. In addition, we have the following behaviour for small $u$ and $v$ respectively
\begin{equation}
H^{(m)}_{\tau}(u,v) \sim u^{\frac{\tau}{2}},~~~~~H^{(m)}_{\tau}(u,v) \sim \frac{1}{v^{\Delta_\phi-m}}
\end{equation}
Below we will be interested in the case in which $\Delta_\phi-m$ is an integer. In this case we can also get a $\log^2v$ behaviour. As we turn on the coupling, the four-point function becomes
\begin{equation}
{\cal G}(u,v) ={\cal G}^{(0)}(u,v) + g \,{\cal G}^{(1)}(u,v)+\cdots
\end{equation}
where ${\cal G}^{(1)}(u,v)$ admits the decomposition 
\begin{equation}
{\cal G}^{(1)}(u,v) = \sum_{\tau,\rho} \left( A_{\tau,\rho} H^{(\rho)}_{\tau}(u,v)+\log u B_{\tau,\rho} H^{(\rho)}_{\tau}(u,v) \right)
\end{equation}
where $\tau$ runs over the spectrum of twists at the $g=0$ point and $\rho$ turns out to run over the spectrum of twists plus integers. Now we make the following powerful observation. The functions $H^{(\rho)}_{\tau}(u,v)$ have a well understood/computable expansion around $u,v=0$. The form of this expansion is such that crossing symmetry can be solved order by order! becoming an algebraic problem. Let us analyse some examples in detail.

\subsection{Example}
Let us come back to the example of generalised free fields in two dimensions. As mentioned before, for large enough $\Delta_\phi$ and to all orders in $(1-\bar z)$ the 2d TCB are of the form
\begin{equation}
H^{(0)}_{\tau}(u,v) =c_\tau  \left( \frac{\bar z}{1-\bar z}\right)^{\Delta_\phi} z^{\tau/2} F_\frac{\tau}{2}(z)
\end{equation}
Plugging this expression into the recurrence relation (\ref{recurrence}) we see that the functions $H^{(m)}_{\tau}(u,v)$ have the factorised form
\begin{equation}
H^{(m)}_{\tau}(u,v) =c_\tau  \bar H^{(m)}_{\tau}(\bar z) z^{\tau/2} F_\frac{\tau}{2}(z)
\end{equation}
with 
\begin{equation}
\label{recurrence2d}
 \bar {D} \bar H^{(m+1)}_{\tau}(\bar z)=\bar H^{(m)}_{\tau}(\bar z),~~~H^{(0)}_{\tau}(\bar z)= \left( \frac{\bar z}{1-\bar z}\right)^{\Delta_\phi}
\end{equation}
and the boundary condition $\bar H^{(m)}_{\tau}(\bar z) \sim (1-\bar z)^{-(\Delta_\phi-m)}$. This allows to find $\bar H^{(m)}_{\tau}(\bar z)$ easily as an expansion in $(1-\bar z)$. For the first few cases this expansion can be re-summed. For instance

\begin{equation}
\bar H^{(1)}_{\tau}(\bar z)=c_\tau \frac{\bar z^{\Delta_\phi}}{(1-\bar z)^{\Delta_\phi-1}} \frac{~_2F_1(\Delta_\phi,\Delta_\phi,1+\Delta_\phi;\bar z)}{\Delta_\phi(\Delta_\phi-1)}
\end{equation}
Note that in the problem (\ref{recurrence2d}) the dependence on the twist $\tau$ has completely dropped out. As a result, the functions  $H^{(m)}_{\tau}(z,\bar z)$ in two dimensions have the following factorised form
\begin{equation}
H^{(m)}_{\tau}(z,\bar z) =c_\tau  \bar H^{(m)}(\bar z) z^{\tau/2} F_\frac{\tau}{2}(z)
\end{equation}
\subsubsection{Integer $\Delta_\phi$}
A nice structure arises for integer but not necessarily large $\Delta_\phi$ . As before, the divergent part of the TCB is captured by
\begin{equation}
H^{(0)}_{\tau}(z,\bar z) =c_\tau  \left( \frac{\bar z}{1-\bar z}\right)^{\Delta_\phi} z^{\tau/2} F_\frac{\tau}{2}(z)
\end{equation}
where $\tau=2\Delta_\phi + 2n$ for double trace operators. Now let us construct explicitly the functions $H^{(m)}_{\tau}(z,\bar z)$. In doing so, we will keep only the pieces with enhanced divergence respect to a single conformal block. To be more precise, we will keep contributions such that enhanced divergences (with respect to a single conformal block) as $\bar z \to 1$ are generated upon the application of the Casimir operator ${\cal C}$ a finite number of times. These contributions may be a negative power of $(1-\bar z)$ or a piece like $(1-\bar z)^p\log^2(1-\bar z) $ for any $p$. As already mentioned, $H^{(m)}_{\tau}(z,\bar z)$ has a factorised form in two dimensions, so that we will only deal with $\bar H^{(m)}(\bar z)$. From the recursion relations (\ref{recurrence2d}), let us compute the sequence of functions $\bar H^{(m)}(\bar z)$ for different values of $\Delta_\phi$. For instance
\begin{center}
\begin{tabular}{ | l |l | }
\hline
   $\Delta_\phi=2$ & $\Delta_\phi=3$\\
  \hline 	
  $ \bar H^{(0)}(\bar z) = \left( \frac{\bar z}{1-\bar z}\right)^2$ & $\bar H^{(0)}(\bar z) = \left( \frac{\bar z}{1-\bar z}\right)^{3} $ \\
  $\bar H^{(1)}(\bar z) = \frac{1}{1-\bar z}$ & $\bar H^{(1)}(\bar z) = \frac{4\bar z-3}{4(1-\bar z)^2}$  \\
  $\bar H^{(2)}(\bar z) = \frac{1}{2}\log^2(1-\bar z) $ & $\bar H^{(2)}(\bar z) = \frac{1}{4(1-\bar z)} -\frac{1}{4}\log^2(1-\bar z) $  \\
  \hline  
\end{tabular}
\end{center}
and so on. The general structure is as follows. $\bar H^{(m)}(\bar z)$ contains power law divergent terms for $m=0,\cdots,\Delta_\phi-1$. $\bar H^{(m)}(\bar z)$ contains $\log^2(1-\bar z)$ terms for $m>1$, and for $m \geq \Delta_\phi$  it is of the form $\bar H^{(m)}(\bar z) = g_m(\bar z)\log^2(1-\bar z)$ with $g_m(\bar z) \sim (1-\bar z)^{m-\Delta_\phi}$. 

Let us understand the consequences of this for the spectrum of the theory at first order in $g$. First let us assume we have only double trace operators in the OPE $\phi \times \phi$. At first order in $g$
\begin{equation}
{\cal G}^{(1)}(z,\bar z) = 1+\sum_{\tau,\rho} \left( A_{\tau,\rho} H^{(\rho)}_{\tau}(z,\bar z)+\log z \bar z B_{\tau,\rho} H^{(\rho)}_{\tau}(z,\bar z) \right)
\end{equation}
It is convenient to write the crossing equation as
\begin{equation}
\left(\frac{1-z}{z}\right)^{\Delta_\phi} {\cal G}^{(1)}(z,\bar z) =\left(\frac{\bar z}{1-\bar z} \right)^{\Delta_\phi} {\cal G}^{(1)}(1-\bar z,1-z),
\end{equation}
where we have chosen crossing to act as $z \leftrightarrow 1-\bar z$. We now make the following simple observation. Since all intermediate operators have twist $\tau=2\Delta_\phi+2n$, all terms on the r.h.s. behave as  $(1-\bar z)^{-\Delta_\phi}(1-\bar z)^{\Delta_\phi+n}$ as $\bar z \to 1$. As a result, the r.h.s. does not have power law divergences at $\bar z =1$. Given the behaviour of $\bar H^{(m)}(\bar z)$ around $\bar z = 1$ we see that all functions $\bar H^{(m)}(\bar z)$  with $m=0,1,\cdots, \Delta_\phi-1$ are forbidden. If they were present on the l.h.s. they would produce a divergence not present on the r.h.s. On the other hand, functions with higher $m$ are also forbidden, since they would lead to terms containing $\log^2(1-\bar z)$, which cannot be produced on the r.h.s. at one loop. We then arrive to the following remarkable conclusion: at first order in $g$ for the present models we are allowed to have only solutions with finite support in the spin (or which decay faster than any power!).  A similar argument can be carried out also in $d=4$, with the same conclusions. This justifies, for instance, some of the claims made in \cite{Heemskerk:2009pn}. 

Next, let us consider a more interesting situation. Imagine that $\phi$ itself is present in the OPE $\phi \times \phi$ at order $g$. In this case ${\cal G}^{(1)}(z,\bar z)$ contains the following piece:
\begin{eqnarray}
{\cal G}^{(1)}(z,\bar z) &=& a_\phi G_{\Delta_\phi,0}(u,v) \\
&=& a_\phi  (z \bar z)^{\Delta_\phi/2} F_\frac{\Delta_\phi}{2}(z)F_\frac{\Delta_\phi}{2}(\bar z) \nonumber
\end{eqnarray}
where $a_\phi$ is the (square) OPE coefficient with which $\phi$ appears. This term acts as a source in the crossing equations. Now
\begin{eqnarray}
\label{crossingdiv}
&\left(\frac{1-z}{z}\right)^{\Delta_\phi}  \sum_{\tau,\rho} \left.\left( A_{\tau,\rho} H^{(\rho)}_{\tau}(z,\bar z)+\log z \bar z B_{\tau,\rho} H^{(\rho)}_{\tau}(z,\bar z) \right) \right|_{div} = \nonumber\\
&a_\phi  \bar z^{\Delta_\phi} \left(\frac{1-z}{1-\bar z} \right)^{\Delta_\phi/2}  F_\frac{\Delta_\phi}{2}(1-z) F_\frac{\Delta_\phi}{2}(1-\bar z)
\end{eqnarray}
The sum on the l.h.s. of (\ref{crossingdiv}) has to reproduce the divergence on the r.h.s. This implies the sum over $\rho$ starts at $\rho=\Delta_\phi/2$ and is such that the precise power law divergence is reproduced. Also, terms of the form $\log^2(1-\bar z)$ should be absent. Furthermore, for each value of $\rho$ the sum over $\tau$ should be such that the behaviour in $z$ is reproduced to all orders. For instance, if we are interested in the anomalous dimensions then we can use
\begin{equation}
F_\frac{\Delta_\phi}{2}(1-z) = -\frac{\Gamma(\Delta_\phi)}{\Gamma^2(\Delta_\phi/2)}~_2F_1(\frac{\Delta_\phi}{2},\frac{\Delta_\phi}{2},1; z) \log z
\end{equation}
up to an holomorphic function at $z=0$. This leads to the following equation
\begin{widetext}
\begin{eqnarray}
\label{TC}
\sum_{\tau,\rho} B_{\tau,\rho} c_\tau H^{(\rho)}(\bar z) z^{\tau/2} F_{\tau/2}(z) =- a_\phi \frac{(z \bar z)^{\Delta_\phi}}{((1-z)(1-\bar z))^{\Delta_\phi/2}}\frac{\Gamma(\Delta_\phi)}{\Gamma^2(\Delta_\phi/2)}F_{\Delta_\phi/2}(1-\bar z) ~_2F_1(\frac{\Delta_\phi}{2},\frac{\Delta_\phi}{2},1; z)
\end{eqnarray}
\end{widetext}
Note that the crossing equation has become completely algebraic as both sides have good convergence properties around $z=0,\bar z=1$! This equation is general. Let us solve (\ref{TC}) in some examples. 
\subsubsection*{$\Delta_\phi=2$}
In this case (\ref{TC}) becomes
\begin{equation}
\label{TCdelta2}
\sum_{\tau,\rho} B_{\tau,\rho} c_\tau \bar H^{(\rho)}(\bar z) z^{\tau/2} F_{\tau/2}(z) =-  \frac{a_\phi z^2}{(1-z)^2(1-\bar z)}
\end{equation}
The sum over twists runs over $\tau=4+2n$. In order to reproduce the divergence on the r.h.s. the sum over $\rho$ should start at $\rho=1$. In order not to produce $\log^2(1-\bar z)$ the sum over $\rho$ should stop also at $\rho=1$. This implies that the expansion of the anomalous dimensions in inverse powers of the conformal spin has exactly one term! This result is valid to all orders in inverse powers of the spin, and holds for all values of the twist. Setting $\rho=1$ we then obtain
\begin{equation}
\sum_{\tau} B_{\tau,1} c_\tau z^{\tau/2} F_{\tau/2}(z) =- \frac{a_\phi  z^2}{(1-z)^2}
\end{equation}
Which is simply solved by
\begin{equation}
B_{2\Delta_\phi+2n,1} =- a_\phi
\end{equation}
This translates into the following result for the anomalous dimensions of double trace operators at first order in $g$ and to all orders in $1/\ell$
\begin{equation}
\gamma_{n,\ell}= - \frac{2 a_\phi}{(\ell+n+2)(\ell+n+1)}
\end{equation}
\subsubsection*{$\Delta_\phi=4$}
This case is more interesting and, to our knowledge, the results unknown. Equation (\ref{TC}) becomes
\begin{eqnarray}
\label{TCdelta4}
\sum_{\tau,\rho} B_{\tau,\rho} c_\tau \bar H^{(\rho)}(\bar z) && z^{\tau/2} F_{\tau/2}(z) \\
&&=- a_\phi \frac{6 z^4 (1+z)}{(1-z)^5} \left(\frac{1}{(1-\bar z)^2}-\frac{3}{1-\bar z} \right) \nonumber
\end{eqnarray}
An advantage of the factorisation into holomorphic and anti-holomorphic functions is that we can solve the problem in two steps. More precisely $B_{\tau,\rho}$ factorises into a function of $\tau$ times a function of $\rho$. Let us first focus in the $\rho$ dependence. In order to reproduce the correct divergence around $\bar z=1$ the sum over $\rho$ should include $\rho=2$ and $\rho=3$. This in turn will produce a term proportional to $\log^2(1-\bar z)$. In order to cancel this term we must include $\rho=4$ and so on. More precisely
\begin{equation}
\frac{1}{(1-\bar z)^2}-\frac{3}{1-\bar z} = \sum_{\rho=2} \alpha_\rho \bar H^{(\rho)}(\bar z)
\end{equation}
the coefficients $\alpha_\rho$ can be found recursively by applying $\bar D$ repeatedly to both sides of the equation. For instance, applying $\bar D$ twice and using $\bar D^2 \bar H^{(2)}(\bar z) =\bar H^{(0)}(\bar z)$ we can fix $\alpha_2$. Applying $\bar D$ once more we can fix $\alpha_3$, and so on. Note that in order to carry out this procedure we don't need to know the explicit form of the functions $\bar H^{(\rho)}(\bar z)$. A similar procedure works in higher dimensions. Remarkably, the coefficients $\alpha_\rho$ have the following closed form expression
\begin{equation}
\alpha_\rho= 2^{\rho-1}(5 \times 3^\rho-9)
\end{equation}
Let us now turn to the dependence on $n$. The problem we need to solve is
\begin{equation}
\sum_{\tau=2\Delta_\phi+2n} B_{\tau,\rho} c_\tau z^{\tau/2} F_{\tau/2}(z) =- a_\phi \frac{6 z^4 (1+z)}{(1-z)^5} 
\end{equation}
Which is solved by
\begin{equation}
B_{2\Delta_\phi+2n,\rho}  =- \frac{3 a_\phi}{4}(n^2+7n+8) 
\end{equation}
This leads to the following prediction for the anomalous dimension of double trace operators in this case
\begin{equation}
\gamma_{n,\ell}=- 54 a_\phi(n^2+7n+8)\frac{J^2-1}{J^2(J^2-2)(J^2-6)},
\end{equation}
where $J^2=(\ell+4+n)(\ell+3+n)$. This prediction is valid to all orders in $1/\ell$.

\section{Comparison to HS conformal blocks}

For theories with HS symmetry we can consider the decomposition of a four-point correlator in HS conformal blocks. These should be labeled by irreducible representations of the HS algebra and should re-sum the contribution of the entire HS multiplet. While we don't know how to do this in general, let us do it for the following interesting example. In the theory of a free scalar field $\varphi$ of dimension $\Delta_{\varphi}=\frac{d-2}{2}$ consider the four point correlator
\begin{equation}
\langle {\cal O}_p(1){\cal O}_p(2){\cal O}_p(3){\cal O}_p(4) \rangle = \frac{{\cal G}_p(u,v)}{\left(x_{12}^2 x_{34}^2\right)^{p \Delta_\varphi}}
\end{equation}
where ${\cal O}_p = \frac{\varphi^p}{\sqrt{p!}}$. It is straightforward to compute this correlator, although one needs to keep track of symmetry factors. We would like to organize the result as a sum over intermediate states with fixed length $L$. Such tower corresponds to a HS multiplet, whose lowest weight state is $\varphi^L$. In order to isolate this contribution we focus on diagrams with exactly $p-L/2$ contractions between $x_1$ and $x_2$ and $p-L/2$ contractions between $x_3$ and $x_4$. Each contribution corresponds to a HS conformal block and has the form
\begin{equation}
{\cal H}_L(u,v) = \left(\frac{u}{v}\right)^{\frac{L}{2}} ~_2F_1(-L/2,-L/2,1;v) 
\end{equation}
where we have set $d=4$ for simplicity. The four-point function can then be written as 
\begin{equation}
{\cal G}_n(u,v)=\sum_{L=0,2,\cdots}^{2p} a_L {\cal H}_L(u,v) 
\end{equation}
where $a_L$ is proportional to the square of the OPE coefficient of the lowest weight state and $a_0=1$. Consider now the crossing relations:
\begin{equation}
{\cal G}_p(u,v)={\cal G}_p(u/v,1/v) ,~~~v^p {\cal G}_p(u,v) =u^p {\cal G}_p(v,u)
\end{equation}
The first relation is automatically satisfied by each HS conformal block. On the other hand, given the finite sum over HS conformal blocks for each $p$, the second relation fixes all relative coefficients $a_L$! this is reminiscent of what happens for Virasoro conformal blocks in minimal models.  

Let us compare the HS conformal blocks to the twist conformal blocks defined above. $L=0$ corresponds to the identity operator and in this case the two blocks trivially coincide. $L=2$ corresponds to the HS conserved currents of form $\varphi \partial_{\mu_1} \cdots \partial_{\mu_\ell} \varphi$. In $d=4$ these operators have twist two and hence the two blocks again coincide. These two statements are general, and not only valid for the case at hand. For $L=4$ and higher, the situation is more interesting, since in this case the intermediate operators have twist $\tau=L,L+2,\cdots$. In summary
\begin{eqnarray}
{\cal H}_0(u,v) &=&H_0(u,v) \\
{\cal H}_2(u,v) &=&H_2(u,v)\\
{\cal H}_L(u,v) &=& H_L(u,v) + \sum_{m=1} c_{L}^m H_{L+2m}(u,v) 
\end{eqnarray}
The coefficients  $c_{L}^m$ can be uniquely fixed by requiring the cancelation of sub-leading powers in $u$.  
 
\section{Outlook}
In this letter we have proposed a new method to study CFT around points of large twist degeneracy. This method transforms the crossing equations into an algebraic problem and allows to solve the theory perturbatively around large spin. The method has a wide range of applicability, does not rely on a Lagrangian description and is valid for arbitrary dimensions. As an example we computed the anomalous dimensions for scalar models around generalised free fields, in two dimensions. For $\Delta_\phi=2$ and at first order in the breaking parameter $g$, our method offers a simple explanation of why the expansions in inverse powers of the conformal spin truncate after a single term. Although we have shown how this works in detail for $d=2$, the method and the result are easily generalised to higher dimensions. For instance, this truncation also holds for the $O(N)$ model in $d=4-\epsilon$, see {\it e.g.} explicit results in \cite{Lang:1992zw,Giombi:2016hkj}, and again, our method explains the reason! For theories with HS symmetry, twist conformal blocks are connected to HS conformal blocks. Twist conformal blocks appear to be better suited for doing perturbations around large spin.  

Let us mention an important point. In our derivation we have got some milage by assuming the singular behaviour (\ref{smallv}). This assumption was motivated by generalised free fields, but this behaviour is not always true. Note however, that the correct behaviour can be inferred once we select a specific CFT and then it is straightforward to apply the machinery developed in this letter.   

Although the features of the method have been shown in simple examples, the range of applicability is much wider. Let us mention some possible applications.  

\bigskip

\noindent
{\bf CFT in various dimensions.} Although we have presented closed form expressions for $d=2$, we have shown how to systematically construct (as series expansions) the functions $H_\tau^{m}(u,v)$ in any number of dimensions. Furthermore, the behaviour around $u,v \sim 0$ is universal and defined by the theory at $g=0$, so that the method can be readily applied to CFT in general dimensions. It would be interesting to obtain explicit results for specific CFT's and compare them to available results. 

\bigskip

\noindent
{\bf Higher orders in $g$.} The applicability of the method is not restricted to first order in the perturbation parameter $g$. Actually, already interesting things can be said at higher orders in the breaking parameter. Given a first order anomalous dimension $\gamma = g \gamma_1 + \cdots$, the expansion of $u^{\gamma/2}$ in the CPW decomposition will give rise to terms of the form $g^2 \gamma_1^2 \log^2u$ at second order. Under crossing this maps to $g^2 \gamma_1^2 \log^2v$. The CFT data at second order should reproduce this behaviour and crossing will again be an algebraic problem. This is used in \cite{QG} in order to compute $1/N^4$ corrections to anomalous dimensions in large $N$ CFT, in various dimensions. Furthermore, at order $g^3$ we have terms of the form $g^3 \gamma_1^3 \log^3v$. Given the behaviour of the functions $H_\tau^{m}(u,v)$, this behaviour can only be reproduced with a logarithmic behaviour for the anomalous dimensions! {\it e.g.} of the form $\frac{ \log \ell}{\ell^n}$. Furthermore, even in the non-perturbative regime, note that the method proposed here generalises to arbitrary twist ($\gamma_{n,\ell}$ as opposed to $\gamma_{0,\ell}$) the method developed in \cite{Alday:2015ewa}. Having control on the dependence on $n$, one can study several questions. 
\bigskip

\noindent
{\bf Conformal gauge theories.}  We can also study weakly coupled conformal gauge theories. These contain single trace operators whose anomalous dimension for $g \neq 0$ grows logarithmically with the spin. In this case one would need to extend the family of functions studied here to include insertions like $\log J$. Alternatively, this can also be studied by inserting $1/J^{2m}$, as done here, making an analytic continuation in $m$, and then taking derivatives with respect to this parameter. Again we will obtain algebraic equations. The proposal of this letter gives a gauge invariant and on-shell method to study weakly coupled gauge theories. In case of theories with a bulk dual, it would be fascinating to use our method to understand the higher spin gauge theory living in the bulk, presumably by exploiting better the relation to HS conformal blocks.

\section*{Acknowledgements} 
We are grateful to O. Aharony, A. Bissi, Z. Komargodski, J. Maldacena and E. Perlmutter for useful comments and discussions. 
This work was supported by ERC STG grant 306260. The author is a Wolfson Royal Society Research Merit Award holder.

\end{document}